\documentclass[preprint,prd,tightenlines,superscriptaddress]{revtex4-1}

\usepackage{graphicx}
\usepackage{dcolumn} 
\usepackage{colordvi}
\usepackage{color}
\usepackage{epstopdf}
\usepackage{amssymb}
\usepackage{amsmath}
\usepackage{url}
\graphicspath{{ps}}
\usepackage{hyperref}
\usepackage{tabularx}
\usepackage[columnwise,pagewise]{lineno}
\usepackage{siunitx}

\newcommand{\ifb}{\ensuremath{{\rm fb}^{-1}}\xspace}

\usepackage{pgf}
\usepackage{pgffor}
\usepackage{pgfplots}
\usepackage{tikz}
\usepackage{units}
\usepackage[super]{nth}
\usepackage{hepnames}

\newcommand{\bdlnu}{\ensuremath{B \to D \, \ell^+\,\nu_{\ell}}\xspace}
\newcommand{\bdslnu}{\ensuremath{B \to D^* \, \ell^+\,\nu_{\ell}}\xspace}
\newcommand{\bddslnu}{\ensuremath{B \to D^{**} \, \ell^+\,\nu_{\ell}}\xspace}

\newcommand{\bxlnu}{\ensuremath{B \to X \, \ell^+\, \nu_{\ell}}\xspace}
\newcommand{\bulnu}{\ensuremath{B \to X_u \, \ell^+\, \nu_{\ell}}\xspace}
\newcommand{\bclnu}{\ensuremath{B \to X_c \, \ell^+\, \nu_{\ell}}\xspace}

\newcommand{\bxenu}{\ensuremath{B \to X \, e^+\, \nu_{e}}\xspace}

\newcommand{\lumi}{\ensuremath{34.6~\ifb}\xspace}
\newcommand{\lumioff}{\ensuremath{3.2~\ifb}\xspace}
\newcommand{\NBB}{\ensuremath{\left(37.7 \pm 0.6 \right) \times 10^6}\xspace}
\newcommand{\chidmeas}{\ensuremath{0.187 \pm 0.010 \text{ (stat.)} \pm 0.019 \text{ (syst.)}}\xspace}

\newcommand{\pee}{\ensuremath{p_{ee}}\xspace}

\usepackage{amssymb}
\usepackage{amsfonts}
\usepackage{upgreek}
\usepackage[tikz]{bclogo}

\usetikzlibrary{pgfplots.colorbrewer,pgfplots.fillbetween}
\usetikzlibrary{shadows.blur}
\usetikzlibrary{shapes.symbols}
\usetikzlibrary{calc,intersections,matrix}
\usetikzlibrary{arrows,fadings,automata,snakes,shapes,shapes.misc,shapes.arrows,trees,positioning,calc,decorations.pathreplacing,arrows.meta}
\tikzfading[name=arrowfading, top color=transparent!0, bottom color=transparent!95]
\tikzset{arrowfill/.style={top color=blue!50, bottom color=blue,}}
\tikzset{>={Latex[width=1.5mm,length=1.5mm]}}
\tikzset{arrowstyle/.style={draw=black,arrowfill, single arrow,minimum height=#1, single arrow,
single arrow head extend=.4cm,}}

\tikzset{
    box/.style={
      rectangle,
	  color=#1,
      draw=black,
      fill=#1,
      thick,
      text=black,
      align=center,
      rounded corners=6pt,
      minimum height=1.5em
    }, 
    hbox/.style={
      rectangle,
      draw=black,
      fill=black,
      thick,
      text=white,
      align=center,
      rounded corners=6pt,
      minimum height=1.5em
    }, 
}
\colorlet{tree0}{blue}
\colorlet{tree100}{white!20!blue}
\colorlet{tree200}{white!40!blue}
\colorlet{tree300}{white!60!blue}
\colorlet{tree400}{white!80!blue}
\colorlet{tree500}{white}
\colorlet{tree600}{white!80!orange}
\colorlet{tree700}{white!60!orange}
\colorlet{tree800}{white!40!orange}
\colorlet{tree900}{white!20!orange}
\colorlet{tree1000}{orange}

\definecolor{Tblue}{HTML}{3465A4}
\definecolor{Tbluedark}{HTML}{204A87}
\definecolor{Tbluelight}{HTML}{729FCF}
\definecolor{Tbluelighter}{HTML}{8CC4FF}	
\definecolor{Tbrown}{HTML}{C17D11}	
\definecolor{Tbrowndark}{HTML}{8F5902}	
\definecolor{Tbrownlight}{HTML}{E9B96E}	
\definecolor{Tgray}{HTML}{888A85}
\definecolor{Tgraydark}{HTML}{555753}	
\definecolor{Tgraydarker}{HTML}{2E3436}	
\definecolor{Tgraylight}{HTML}{BABDB6}	
\definecolor{Tgraylight2}{HTML}{E4E6E2}	
\definecolor{Tgraylight3}{HTML}{F0F2EE}	
\definecolor{Tgreen}{HTML}{73D216}	
\definecolor{Tgreendark}{HTML}{4E9A06}
\definecolor{Tgreenlight}{HTML}{8AE234}	
\definecolor{Tred}{HTML}{CC0000}	
\definecolor{Treddark}{HTML}{A40000}
\definecolor{Tredlight}{HTML}{EF2929}
\definecolor{Tlilac}{HTML}{75507B}
\definecolor{Tlilacdark}{HTML}{5C3566}
\definecolor{Tlilaclight}{HTML}{AD7FA8}
\definecolor{Tyellow}{HTML}{EDD400}	
\definecolor{Tyellowdark}{HTML}{C4A000}	
\definecolor{Tyellowlight}{HTML}{FCE94F}
\definecolor{Torange}{HTML}{F57900}	
\definecolor{Torangedark}{HTML}{CE5C00}
\definecolor{Torangelight}{HTML}{FCAF3E}

\input ./belle2sym.tex

\makeatletter
\let\LN@align\align
\let\LN@endalign\endalign
\renewcommand{\align}{\linenomath\LN@align}
\renewcommand{\endalign}{\LN@endalign\endlinenomath}
\let\LN@gather\gather
\let\LN@endgather\endgather
\renewcommand{\gather}{\linenomath\LN@gather}
\renewcommand{\endgather}{\LN@endgather\endlinenomath}
\makeatother

\begin{document}

\def\belletwo {\it {Belle II}}

\vspace*{-3\baselineskip}
\resizebox{!}{3cm}{\includegraphics{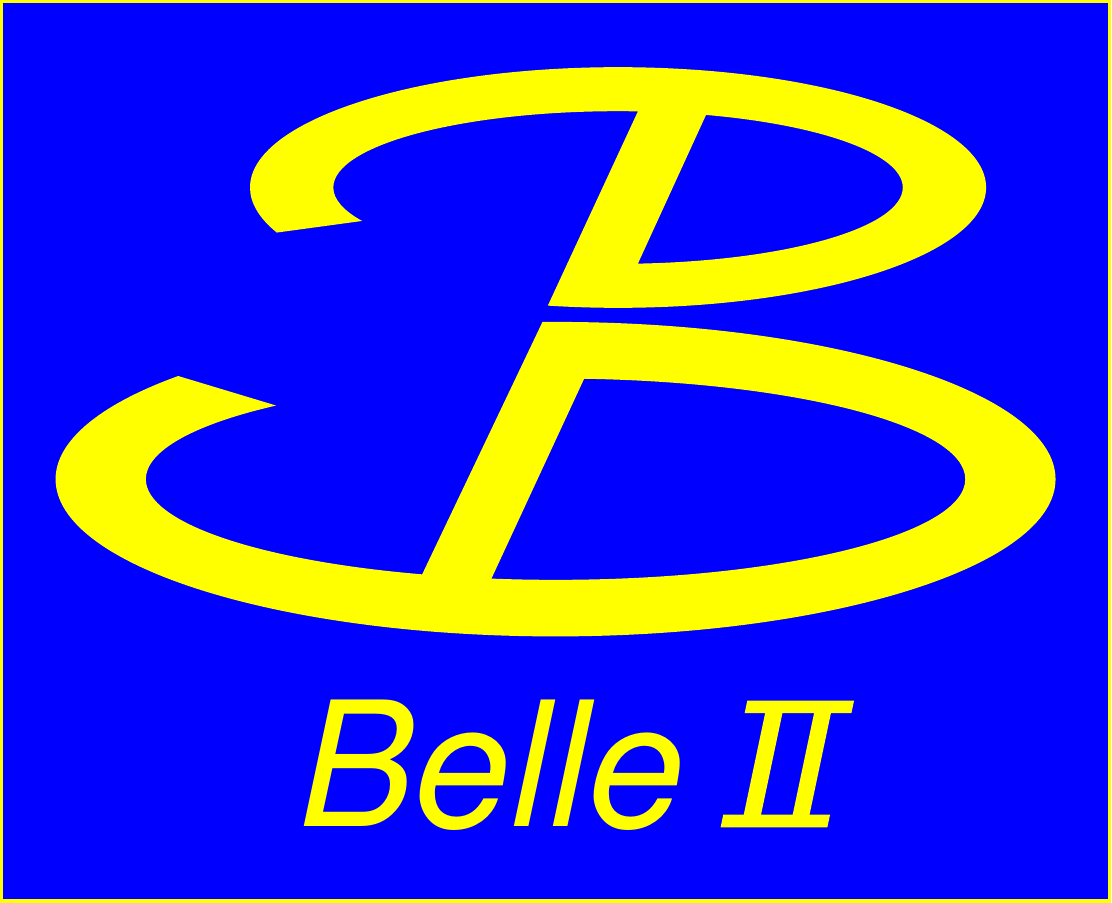}}

\vspace*{-5\baselineskip}
\begin{flushright}
BELLE2-CONF-PH-2021-04\\
\today
\end{flushright}

\title { \quad\\[0.5cm] Measurement of the time-integrated mixing probability $\chi_d$ with a semileptonic double-tagging strategy and \lumi of Belle~II collision data}
\newcommand{\instCPPM}{Aix Marseille Universit\'{e}, CNRS/IN2P3, CPPM, 13288 Marseille, France}
\newcommand{\instBeihang}{Beihang University, Beijing 100191, China}
\newcommand{\instBNL}{Brookhaven National Laboratory, Upton, New York 11973, U.S.A.}
\newcommand{\instBINP}{Budker Institute of Nuclear Physics SB RAS, Novosibirsk 630090, Russian Federation}
\newcommand{\instCMU}{Carnegie Mellon University, Pittsburgh, Pennsylvania 15213, U.S.A.}
\newcommand{\instCinvestavIPN}{Centro de Investigacion y de Estudios Avanzados del Instituto Politecnico Nacional, Mexico City 07360, Mexico}
\newcommand{\instPrague}{Faculty of Mathematics and Physics, Charles University, 121 16 Prague, Czech Republic}
\newcommand{\instChiangMai}{Chiang Mai University, Chiang Mai 50202, Thailand}
\newcommand{\instChiba}{Chiba University, Chiba 263-8522, Japan}
\newcommand{\instChonnam}{Chonnam National University, Gwangju 61186, South Korea}
\newcommand{\instConacyt}{Consejo Nacional de Ciencia y Tecnolog\'{\i}a, Mexico City 03940, Mexico}
\newcommand{\instDESY}{Deutsches Elektronen--Synchrotron, 22607 Hamburg, Germany}
\newcommand{\instDuke}{Duke University, Durham, North Carolina 27708, U.S.A.}
\newcommand{\instITAR}{Institute of Theoretical and Applied Research (ITAR), Duy Tan University, Hanoi 100000, Vietnam}
\newcommand{\instRomaENEA}{ENEA Casaccia, I-00123 Roma, Italy}
\newcommand{\instEri}{Earthquake Research Institute, University of Tokyo, Tokyo 113-0032, Japan}
\newcommand{\instJuelich}{Forschungszentrum J\"{u}lich, 52425 J\"{u}lich, Germany}
\newcommand{\instFuJen}{Department of Physics, Fu Jen Catholic University, Taipei 24205, Taiwan}
\newcommand{\instFudan}{Key Laboratory of Nuclear Physics and Ion-beam Application (MOE) and Institute of Modern Physics, Fudan University, Shanghai 200443, China}
\newcommand{\instGoettingen}{II. Physikalisches Institut, Georg-August-Universit\"{a}t G\"{o}ttingen, 37073 G\"{o}ttingen, Germany}
\newcommand{\instGifu}{Gifu University, Gifu 501-1193, Japan}
\newcommand{\instSOKENDAI}{The Graduate University for Advanced Studies (SOKENDAI), Hayama 240-0193, Japan}
\newcommand{\instGyeongsang}{Gyeongsang National University, Jinju 52828, South Korea}
\newcommand{\instHanyang}{Department of Physics and Institute of Natural Sciences, Hanyang University, Seoul 04763, South Korea}
\newcommand{\instKEK}{High Energy Accelerator Research Organization (KEK), Tsukuba 305-0801, Japan}
\newcommand{\instJPARC}{J-PARC Branch, KEK Theory Center, High Energy Accelerator Research Organization (KEK), Tsukuba 305-0801, Japan}
\newcommand{\instHSE}{National Research University Higher School of Economics, Moscow 101000, Russian Federation}
\newcommand{\instIISER}{Indian Institute of Science Education and Research Mohali, SAS Nagar, 140306, India}
\newcommand{\instIITBhubaneswar}{Indian Institute of Technology Bhubaneswar, Satya Nagar 751007, India}
\newcommand{\instIITGuwahati}{Indian Institute of Technology Guwahati, Assam 781039, India}
\newcommand{\instIITHyderabad}{Indian Institute of Technology Hyderabad, Telangana 502285, India}
\newcommand{\instIITMadras}{Indian Institute of Technology Madras, Chennai 600036, India}
\newcommand{\instIndiana}{Indiana University, Bloomington, Indiana 47408, U.S.A.}
\newcommand{\instIHEPRussia}{Institute for High Energy Physics, Protvino 142281, Russian Federation}
\newcommand{\instHEPHYVienna}{Institute of High Energy Physics, Vienna 1050, Austria}
\newcommand{\instHiroshima}{Hiroshima University, Higashi-Hiroshima, Hiroshima 739-8530, Japan}
\newcommand{\instIHEPChina}{Institute of High Energy Physics, Chinese Academy of Sciences, Beijing 100049, China}
\newcommand{\instIPP}{Institute of Particle Physics (Canada), Victoria, British Columbia V8W 2Y2, Canada}
\newcommand{\instIOP}{Institute of Physics, Vietnam Academy of Science and Technology (VAST), Hanoi, Vietnam}
\newcommand{\instIFIC}{Instituto de Fisica Corpuscular, Paterna 46980, Spain}
\newcommand{\instFrascati}{INFN Laboratori Nazionali di Frascati, I-00044 Frascati, Italy}
\newcommand{\instNapoliINFN}{INFN Sezione di Napoli, I-80126 Napoli, Italy}
\newcommand{\instPadovaINFN}{INFN Sezione di Padova, I-35131 Padova, Italy}
\newcommand{\instPerugiaINFN}{INFN Sezione di Perugia, I-06123 Perugia, Italy}
\newcommand{\instPisaINFN}{INFN Sezione di Pisa, I-56127 Pisa, Italy}
\newcommand{\instRomaINFN}{INFN Sezione di Roma, I-00185 Roma, Italy}
\newcommand{\instRomaTreINFN}{INFN Sezione di Roma Tre, I-00146 Roma, Italy}
\newcommand{\instTorinoINFN}{INFN Sezione di Torino, I-10125 Torino, Italy}
\newcommand{\instTriesteINFN}{INFN Sezione di Trieste, I-34127 Trieste, Italy}
\newcommand{\instJAEA}{Advanced Science Research Center, Japan Atomic Energy Agency, Naka 319-1195, Japan}
\newcommand{\instMainz}{Johannes Gutenberg-Universit\"{a}t Mainz, Institut f\"{u}r Kernphysik, D-55099 Mainz, Germany}
\newcommand{\instGiessen}{Justus-Liebig-Universit\"{a}t Gie\ss{}en, 35392 Gie\ss{}en, Germany}
\newcommand{\instKarlsruhe}{Institut f\"{u}r Experimentelle Teilchenphysik, Karlsruher Institut f\"{u}r Technologie, 76131 Karlsruhe, Germany}
\newcommand{\instISU}{Iowa State University, Ames, Iowa 50011, U.S.A.}
\newcommand{\instKitasato}{Kitasato University, Sagamihara 252-0373, Japan}
\newcommand{\instKISTI}{Korea Institute of Science and Technology Information, Daejeon 34141, South Korea}
\newcommand{\instKoreaUnivKU}{Korea University, Seoul 02841, South Korea}
\newcommand{\instKSU}{Kyoto Sangyo University, Kyoto 603-8555, Japan}
\newcommand{\instKyungpook}{Kyungpook National University, Daegu 41566, South Korea}
\newcommand{\instLPI}{P.N. Lebedev Physical Institute of the Russian Academy of Sciences, Moscow 119991, Russian Federation}
\newcommand{\instLNNU}{Liaoning Normal University, Dalian 116029, China}
\newcommand{\instLMU}{Ludwig Maximilians University, 80539 Munich, Germany}
\newcommand{\instLuther}{Luther College, Decorah, Iowa 52101, U.S.A.}
\newcommand{\instMNITJaipur}{Malaviya National Institute of Technology Jaipur, Jaipur 302017, India}
\newcommand{\instMPP}{Max-Planck-Institut f\"{u}r Physik, 80805 M\"{u}nchen, Germany}
\newcommand{\instMPGHLL}{Semiconductor Laboratory of the Max Planck Society, 81739 M\"{u}nchen, Germany}
\newcommand{\instMcGill}{McGill University, Montr\'{e}al, Qu\'{e}bec, H3A 2T8, Canada}
\newcommand{\instMEPhI}{Moscow Physical Engineering Institute, Moscow 115409, Russian Federation}
\newcommand{\instNagoya}{Graduate School of Science, Nagoya University, Nagoya 464-8602, Japan}
\newcommand{\instNagoyaIAR}{Institute for Advanced Research, Nagoya University, Nagoya 464-8602, Japan}
\newcommand{\instNagoyaKMI}{Kobayashi-Maskawa Institute, Nagoya University, Nagoya 464-8602, Japan}
\newcommand{\instNaraWu}{Nara Women's University, Nara 630-8506, Japan}
\newcommand{\instNTUTaiwan}{Department of Physics, National Taiwan University, Taipei 10617, Taiwan}
\newcommand{\instNUUTaiwan}{National United University, Miao Li 36003, Taiwan}
\newcommand{\instKrakow}{H. Niewodniczanski Institute of Nuclear Physics, Krakow 31-342, Poland}
\newcommand{\instNiigata}{Niigata University, Niigata 950-2181, Japan}
\newcommand{\instNSU}{Novosibirsk State University, Novosibirsk 630090, Russian Federation}
\newcommand{\instOkinawa}{Okinawa Institute of Science and Technology, Okinawa 904-0495, Japan}
\newcommand{\instOsakaCity}{Osaka City University, Osaka 558-8585, Japan}
\newcommand{\instRCNP}{Research Center for Nuclear Physics, Osaka University, Osaka 567-0047, Japan}
\newcommand{\instPNNL}{Pacific Northwest National Laboratory, Richland, Washington 99352, U.S.A.}
\newcommand{\instPanjab}{Panjab University, Chandigarh 160014, India}
\newcommand{\instPanjabPAU}{Punjab Agricultural University, Ludhiana 141004, India}
\newcommand{\instRIKENMSL}{Meson Science Laboratory, Cluster for Pioneering Research, RIKEN, Saitama 351-0198, Japan}
\newcommand{\instSeoul}{Seoul National University, Seoul 08826, South Korea}
\newcommand{\instSPU}{Showa Pharmaceutical University, Tokyo 194-8543, Japan}
\newcommand{\instSoochow}{Soochow University, Suzhou 215006, China}
\newcommand{\instSoongsil}{Soongsil University, Seoul 06978, South Korea}
\newcommand{\instLjubljanaJSI}{J. Stefan Institute, 1000 Ljubljana, Slovenia}
\newcommand{\instKyiv}{Taras Shevchenko National Univ. of Kiev, Kiev, Ukraine}
\newcommand{\instTata}{Tata Institute of Fundamental Research, Mumbai 400005, India}
\newcommand{\instTUM}{Department of Physics, Technische Universit\"{a}t M\"{u}nchen, 85748 Garching, Germany}
\newcommand{\instTelAviv}{Tel Aviv University, School of Physics and Astronomy, Tel Aviv, 69978, Israel}
\newcommand{\instToho}{Toho University, Funabashi 274-8510, Japan}
\newcommand{\instTohoku}{Department of Physics, Tohoku University, Sendai 980-8578, Japan}
\newcommand{\instTitech}{Tokyo Institute of Technology, Tokyo 152-8550, Japan}
\newcommand{\instTokyoMetropolitan}{Tokyo Metropolitan University, Tokyo 192-0397, Japan}
\newcommand{\instUAS}{Universidad Autonoma de Sinaloa, Sinaloa 80000, Mexico}
\newcommand{\instNapoliUNIV}{Dipartimento di Scienze Fisiche, Universit\`{a} di Napoli Federico II, I-80126 Napoli, Italy}
\newcommand{\instPadovaUNIV}{Dipartimento di Fisica e Astronomia, Universit\`{a} di Padova, I-35131 Padova, Italy}
\newcommand{\instPerugiaUNIV}{Dipartimento di Fisica, Universit\`{a} di Perugia, I-06123 Perugia, Italy}
\newcommand{\instPisaUNIV}{Dipartimento di Fisica, Universit\`{a} di Pisa, I-56127 Pisa, Italy}
\newcommand{\instRomaTreUNIV}{Dipartimento di Matematica e Fisica, Universit\`{a} di Roma Tre, I-00146 Roma, Italy}
\newcommand{\instTorinoUNIV}{Dipartimento di Fisica, Universit\`{a} di Torino, I-10125 Torino, Italy}
\newcommand{\instTriesteUNIV}{Dipartimento di Fisica, Universit\`{a} di Trieste, I-34127 Trieste, Italy}
\newcommand{\instMontreal}{Universit\'{e} de Montr\'{e}al, Physique des Particules, Montr\'{e}al, Qu\'{e}bec, H3C 3J7, Canada}
\newcommand{\instIJCLab}{Universit\'{e} Paris-Saclay, CNRS/IN2P3, IJCLab, 91405 Orsay, France}
\newcommand{\instIPHC}{Universit\'{e} de Strasbourg, CNRS, IPHC, UMR 7178, 67037 Strasbourg, France}
\newcommand{\instAdelaide}{Department of Physics, University of Adelaide, Adelaide, South Australia 5005, Australia}
\newcommand{\instBonn}{University of Bonn, 53115 Bonn, Germany}
\newcommand{\instUBC}{University of British Columbia, Vancouver, British Columbia, V6T 1Z1, Canada}
\newcommand{\instCincinnati}{University of Cincinnati, Cincinnati, Ohio 45221, U.S.A.}
\newcommand{\instFlorida}{University of Florida, Gainesville, Florida 32611, U.S.A.}
\newcommand{\instHawaii}{University of Hawaii, Honolulu, Hawaii 96822, U.S.A.}
\newcommand{\instHeidelberg}{University of Heidelberg, 68131 Mannheim, Germany}
\newcommand{\instLjubljanaUniLJ}{Faculty of Mathematics and Physics, University of Ljubljana, 1000 Ljubljana, Slovenia}
\newcommand{\instLouisville}{University of Louisville, Louisville, Kentucky 40292, U.S.A.}
\newcommand{\instMalaya}{National Centre for Particle Physics, University Malaya, 50603 Kuala Lumpur, Malaysia}
\newcommand{\instLjubljanaUM}{Faculty of Chemistry and Chemical Engineering, University of Maribor, 2000 Maribor, Slovenia}
\newcommand{\instMelbourne}{School of Physics, University of Melbourne, Victoria 3010, Australia}
\newcommand{\instMississippi}{University of Mississippi, University, Mississippi 38677, U.S.A.}
\newcommand{\instUOM}{University of Miyazaki, Miyazaki 889-2192, Japan}
\newcommand{\instPittsburgh}{University of Pittsburgh, Pittsburgh, Pennsylvania 15260, U.S.A.}
\newcommand{\instUSTC}{University of Science and Technology of China, Hefei 230026, China}
\newcommand{\instSAlabama}{University of South Alabama, Mobile, Alabama 36688, U.S.A.}
\newcommand{\instSCarolina}{University of South Carolina, Columbia, South Carolina 29208, U.S.A.}
\newcommand{\instSydney}{School of Physics, University of Sydney, New South Wales 2006, Australia}
\newcommand{\instUTokyo}{Department of Physics, University of Tokyo, Tokyo 113-0033, Japan}
\newcommand{\instIPMU}{Kavli Institute for the Physics and Mathematics of the Universe (WPI), University of Tokyo, Kashiwa 277-8583, Japan}
\newcommand{\instVictoria}{University of Victoria, Victoria, British Columbia, V8W 3P6, Canada}
\newcommand{\instVPI}{Virginia Polytechnic Institute and State University, Blacksburg, Virginia 24061, U.S.A.}
\newcommand{\instWayneState}{Wayne State University, Detroit, Michigan 48202, U.S.A.}
\newcommand{\instYamagata}{Yamagata University, Yamagata 990-8560, Japan}
\newcommand{\instYerevan}{Alikhanyan National Science Laboratory, Yerevan 0036, Armenia}
\newcommand{\instYonsei}{Yonsei University, Seoul 03722, South Korea}
\newcommand{\instZZU}{Zhengzhou University, Zhengzhou 450001, China}
\affiliation{\instCPPM}
\affiliation{\instBeihang}
\affiliation{\instBNL}
\affiliation{\instBINP}
\affiliation{\instCMU}
\affiliation{\instCinvestavIPN}
\affiliation{\instPrague}
\affiliation{\instChiangMai}
\affiliation{\instChiba}
\affiliation{\instChonnam}
\affiliation{\instConacyt}
\affiliation{\instDESY}
\affiliation{\instDuke}
\affiliation{\instITAR}
\affiliation{\instRomaENEA}
\affiliation{\instEri}
\affiliation{\instJuelich}
\affiliation{\instFuJen}
\affiliation{\instFudan}
\affiliation{\instGoettingen}
\affiliation{\instGifu}
\affiliation{\instSOKENDAI}
\affiliation{\instGyeongsang}
\affiliation{\instHanyang}
\affiliation{\instKEK}
\affiliation{\instJPARC}
\affiliation{\instHSE}
\affiliation{\instIISER}
\affiliation{\instIITBhubaneswar}
\affiliation{\instIITGuwahati}
\affiliation{\instIITHyderabad}
\affiliation{\instIITMadras}
\affiliation{\instIndiana}
\affiliation{\instIHEPRussia}
\affiliation{\instHEPHYVienna}
\affiliation{\instHiroshima}
\affiliation{\instIHEPChina}
\affiliation{\instIPP}
\affiliation{\instIOP}
\affiliation{\instIFIC}
\affiliation{\instFrascati}
\affiliation{\instNapoliINFN}
\affiliation{\instPadovaINFN}
\affiliation{\instPerugiaINFN}
\affiliation{\instPisaINFN}
\affiliation{\instRomaINFN}
\affiliation{\instRomaTreINFN}
\affiliation{\instTorinoINFN}
\affiliation{\instTriesteINFN}
\affiliation{\instJAEA}
\affiliation{\instMainz}
\affiliation{\instGiessen}
\affiliation{\instKarlsruhe}
\affiliation{\instISU}
\affiliation{\instKitasato}
\affiliation{\instKISTI}
\affiliation{\instKoreaUnivKU}
\affiliation{\instKSU}
\affiliation{\instKyungpook}
\affiliation{\instLPI}
\affiliation{\instLNNU}
\affiliation{\instLMU}
\affiliation{\instLuther}
\affiliation{\instMNITJaipur}
\affiliation{\instMPP}
\affiliation{\instMPGHLL}
\affiliation{\instMcGill}
\affiliation{\instMEPhI}
\affiliation{\instNagoya}
\affiliation{\instNagoyaIAR}
\affiliation{\instNagoyaKMI}
\affiliation{\instNaraWu}
\affiliation{\instNTUTaiwan}
\affiliation{\instNUUTaiwan}
\affiliation{\instKrakow}
\affiliation{\instNiigata}
\affiliation{\instNSU}
\affiliation{\instOkinawa}
\affiliation{\instOsakaCity}
\affiliation{\instRCNP}
\affiliation{\instPNNL}
\affiliation{\instPanjab}
\affiliation{\instPanjabPAU}
\affiliation{\instRIKENMSL}
\affiliation{\instSeoul}
\affiliation{\instSPU}
\affiliation{\instSoochow}
\affiliation{\instSoongsil}
\affiliation{\instLjubljanaJSI}
\affiliation{\instKyiv}
\affiliation{\instTata}
\affiliation{\instTUM}
\affiliation{\instTelAviv}
\affiliation{\instToho}
\affiliation{\instTohoku}
\affiliation{\instTitech}
\affiliation{\instTokyoMetropolitan}
\affiliation{\instUAS}
\affiliation{\instNapoliUNIV}
\affiliation{\instPadovaUNIV}
\affiliation{\instPerugiaUNIV}
\affiliation{\instPisaUNIV}
\affiliation{\instRomaTreUNIV}
\affiliation{\instTorinoUNIV}
\affiliation{\instTriesteUNIV}
\affiliation{\instMontreal}
\affiliation{\instIJCLab}
\affiliation{\instIPHC}
\affiliation{\instAdelaide}
\affiliation{\instBonn}
\affiliation{\instUBC}
\affiliation{\instCincinnati}
\affiliation{\instFlorida}
\affiliation{\instHawaii}
\affiliation{\instHeidelberg}
\affiliation{\instLjubljanaUniLJ}
\affiliation{\instLouisville}
\affiliation{\instMalaya}
\affiliation{\instLjubljanaUM}
\affiliation{\instMelbourne}
\affiliation{\instMississippi}
\affiliation{\instUOM}
\affiliation{\instPittsburgh}
\affiliation{\instUSTC}
\affiliation{\instSAlabama}
\affiliation{\instSCarolina}
\affiliation{\instSydney}
\affiliation{\instUTokyo}
\affiliation{\instIPMU}
\affiliation{\instVictoria}
\affiliation{\instVPI}
\affiliation{\instWayneState}
\affiliation{\instYamagata}
\affiliation{\instYerevan}
\affiliation{\instYonsei}
\affiliation{\instZZU}
  \author{F.~Abudin{\'e}n}\affiliation{\instTriesteINFN} 
  \author{I.~Adachi}\affiliation{\instKEK}\affiliation{\instSOKENDAI} 
  \author{R.~Adak}\affiliation{\instFudan} 
  \author{K.~Adamczyk}\affiliation{\instKrakow} 
  \author{P.~Ahlburg}\affiliation{\instBonn} 
  \author{J.~K.~Ahn}\affiliation{\instKoreaUnivKU} 
  \author{H.~Aihara}\affiliation{\instUTokyo} 
  \author{N.~Akopov}\affiliation{\instYerevan} 
  \author{A.~Aloisio}\affiliation{\instNapoliUNIV}\affiliation{\instNapoliINFN} 
  \author{F.~Ameli}\affiliation{\instRomaINFN} 
% ...

\collaboration{The Belle II Collaboration}
\noaffiliation

\begin{abstract}
   We present the first measurement of the time-integrated mixing probability $\chi_d$ using Belle II data collected at a center-of-mass (CM) energy of 10.58~GeV, corresponding to the mass of the $\Upsilon$(4S) resonance, with an integrated luminosity of \lumi at the SuperKEKB $e^+ e^-$ collider. We reconstruct pairs of B mesons both of which decay to semileptonic final states. Using a novel methodology, we measure $\chi_d = \chidmeas$, which is compatible with existing indirect and direct determinations.
\end{abstract}

\maketitle

{\renewcommand{\thefootnote}{\fnsymbol{footnote}}}
\setcounter{footnote}{0}

\section{Introduction}
The time evolution of the neutral $B^0 - \overline B^0$ meson system has been studied by a number of experiments. The first signal for $B^0 - \overline B^0$  mixing was reported by the ARGUS experiment in Ref.~\cite{Prentice:1987ap} and the observed level of mixing provided first indications for the mass scale of the top quark. 
The two $B^0$ mesons produced in an $\Upsilon$(4S) decay evolve in a coherent $P$-wave state. For this reason the neutral $B$ mesons' flavor in the coherent quantum state are always opposite. Due to the quantum entanglement of the $B$ meson states in the coherent quantum state, they cannot change flavor until one of them has decayed. Therefore, the $B$ mesons' flavor can only be determined relative to each other after one of the mesons decays.
The mixing properties of the $B^0 - \overline B^0$ system can be described by four parameters ($x_d$, $y_d$, $q$, and $p$) and its time evolution is described by the Schr\"odinger equation, which depends on the relative time difference between the two neutral $B$ meson decays. The heavy ($H$) and light ($L$) mass eigenstates of the system are related to the  $B^0$ and $\overline B^0$ flavor eigenstates by:
\begin{align}
 | B_{L/H} \rangle = p |  B_d^0 \rangle \pm q | \overline B_d^0 \rangle \, .
\end{align}
In the absence of $CP$ violation in mixing $|q/p| = 1$, and we can express $x_p$ and $y_p$ as a function of the mass difference, $\Delta m_d  = m_H - m_L$, and the lifetime difference, $\Delta \Gamma_d = \Gamma_L - \Gamma_H$ of the two mass eigenstates:
\begin{align}
 x_d = \Delta m_d / \Gamma_d \, , \qquad y_d = \Delta \Gamma_d / \Gamma_d \, ,
\end{align}
where $\Gamma_d = (\Gamma_L + \Gamma_H)/2$ is the average decay width. Experimentally, both parameters can be constrained by measuring the time-integrated mixing probability, 
\begin{align}
\chi_d = \frac{ \Gamma(B^0 \to \overline B^0) }{ \Gamma(B^0 \to B^0) + \Gamma(B^0 \to \overline B^0) }  = \frac{x_d^2 + y_d^2}{2(x_d^2 + 1)} \, ,
\end{align}
and the value of $y_d$ can be determined in combination with direct measurements of $\Delta m_d$ and the $B$ meson lifetime. The current most precise value of $\chi_d$ is obtained by combining the information from time-independent and time-dependent measurements: $\chi_d^{\mathrm{WA}} = 0.186 \pm 0.001$~\cite{pdg:2020}. The world average obtained using only time-independent measurements has a much larger uncertainty and results in $\chi_d^{\mathrm{WA} \, t\mathrm{-in.}} = 0.182 \pm 0.015$.

Here we provide the first direct determination of $\chi_d$ using a time-independent approach and semileptonic \bxenu decays using data recorded by the Belle II experiment at the SuperKEKB $e^+ e^-$ collider in 2019 and 2020. We analyze an integrated luminosity of \lumi of recorded collision data, corresponding to \NBB of $B$ meson pairs. We identify events in which both $B$ mesons decayed via a semileptonic \bxenu transition. The value of $\chi_d$ is obtained by determining the number of $e^\pm e^\pm$ same-sign  ($N_{\mathrm{SS}}$) and $e^\pm e^\mp$ opposite-sign ($N_{\mathrm{OS}}$) electron pair candidates from \bxenu decays, as the charge of the lepton in a semileptonic $B$ decay directly encodes the flavor of the $B$ meson. Contributions from charged $B$ mesons are subtracted using a correction factor $r_B$ and then the number of opposite- and same-sign events are corrected for selection and acceptance effects to determine the time-integrated mixing probability
\begin{align}
 \chi_d = \frac{ N_{\mathrm{SS}}  }{ N_{\mathrm{SS}} + N_{\mathrm{OS}} \cdot \left(  \epsilon_\textrm{OS} / \epsilon_\textrm{SS} \right)^{-1} } \, \cdot \left( 1 + r_B \right) \, .
\end{align}
Here $\epsilon_\textrm{OS}$ and $\epsilon_\textrm{SS}$ denote the efficiencies for opposite-sign and same-sign signal, respectively, which are obtained from studies on simulated samples and corrected using data-driven methods to account for differences in particle identification and reconstruction efficiencies, with $ \left(\epsilon_\textrm{OS} / \epsilon_\textrm{SS}\right) = 0.92 \pm 0.01 \text{ (stat.)}$. Further, $r_B = f_{+0} \cdot \tau_{+0}^2 = 1.2 \pm 0.1$ with $\tau_{+0} = 1.078 \pm 0.004$ denoting the charged and neutral $B$ meson lifetime ratio and $f_{+0} = \mathcal{B}(\Upsilon(4S) \to B^+ \, B^-) / \mathcal{B}(\Upsilon(4S) \to B^0 \, \overline B^0) = 1.058 \pm 0.024$~\cite{pdg:2020}.

\section{The Belle~II detector and data sample}

The Belle~II detector~\cite{Abe:2010sj, ref:b2tip} operates at the SuperKEKB asymmetric-energy  electron-positron collider~\cite{superkekb}, located at the KEK laboratory in Tsukuba, Japan. The detector consists of several nested detector subsystems arranged around the beam pipe in a cylindrical geometry. The innermost subsystem is the vertex detector, which includes two layers of silicon pixels and four outer layers of silicon strip detectors. Currently, the second pixel layer is installed in only a small part of the solid angle, while the remaining vertex detector layers are fully installed. Most of the tracking volume consists of a small-cell drift chamber filled with a helium and ethane mixture gas. Outside the drift chamber, a Cherenkov-light imaging and time-of-propagation detector provides charged-particle identification in the barrel region. In the forward endcap, this function is provided by a proximity-focusing, ring-imaging Cherenkov detector with an aerogel radiator. At higher radius is an electromagnetic calorimeter, consisting of a barrel and two endcap sections made of CsI(Tl) crystals. A uniform 1.5~T magnetic field is provided by a superconducting solenoid situated outside the calorimeter. The $K_L^0$ and muon identification system consists of multiple layers of scintillators in the endcaps, and two layers of scintillators and multiple layers of resistive plate chambers in the barrel region, located between the magnetic flux-return iron plates.

The data used in this analysis were collected at a center-of-mass (CM) energy of 10.58~GeV, corresponding to the mass of the $\Upsilon$(4S) resonance. The energies of the electron and positron beams are $7\gev$ and $4\gev$, respectively, resulting in a boost of $\beta\gamma = 0.28$ of the CM frame relative to the laboratory frame. In addition, \lumioff of off-resonance collision data, collected $\SI{60}{\MeV}$ below the $\Upsilon$(4S) resonance, is used to model background from $e^+ e^-$ continuum processes, i.e. $e^+ e^- \to u\bar u, d\bar d, s\bar s \text{ and } c\bar c$.

Simulated Monte Carlo (MC) samples of \bxlnu signal and background processes are used to obtain the reconstruction efficiencies and study the key kinematic distributions. These events are generated with EvtGen~\cite{Lange:2001uf} and the branching fractions used are summarized in Table~\ref{tab:bfs}. Inclusive semileptonic  \bxlnu\ decays are dominated by \bdlnu\ and \bdslnu\ transitions. We model the \bdlnu\ decays using the BGL parametrization~\cite{Boyd:1994tt} with form factor central values and uncertainties taken from the fit in Ref.~\cite{Glattauer:2015teq}. For \bdslnu\, we use the BGL implementation proposed in Refs.~\cite{Grinstein:2017nlq,Bigi:2017njr} with form factor central values and uncertainties from the fit to the measurement of Ref.~\cite{Waheed:2018djm}. Both semileptonic modes are normalized to the average branching fraction of Ref.~\cite{Amhis:2019ckw} assuming isospin symmetry.  Semileptonic \bddslnu decays with $D^{**} = \{ D_0^*, D_1^*, D_1, D_2^* \}$ denoting the four orbitally excited charmed mesons are modeled using the heavy-quark-symmetry-based form factors proposed in Refs.~\cite{Bernlochner:2016bci,Bernlochner:2017jxt}. To fill the remaining `gap' between the sum of all measured exclusive modes and the inclusive \bclnu branching fraction, we simulate $B \to D^{(*)} \, \pi \, \pi \, \ell^+ \, \nu_\ell$ and $B \to D^{(*)} \, \eta \, \ell^+ \, \nu_\ell$ decays using a model based on a uniform distribution of all final-state particles in phase-space and add them to the simulated samples. Semileptonic \bulnu decays are modeled as a mixture of specific exclusive modes and non-resonant contributions adapted from the approach described in Ref.~\cite{PhysRevD.41.1496}.

\begin{table}[t!]
\caption{Branching fractions for \bclnu and \bulnu processes that were used to generate simulated samples are listed.}
\label{tab:bfs}
\vspace{1ex}
\begin{tabular}{lcc}
\hline\hline
 $\mathcal{B}$ & $B^+$ & $B^0$ \\
 \hline
 \bclnu & \\
 \quad $B \to D \, \ell^+ \, \nu_\ell$ & $\left(2.41 \pm 0.07\right) \times 10^{-2} $ & $\left(2.24 \pm 0.07\right)\times 10^{-2} $ \\
 \quad $B \to D^* \, \ell^+ \, \nu_\ell$ & $\left(5.50 \pm 0.11 \right)\times 10^{-2} $ &$\left( 5.11 \pm 0.11 \right)\times 10^{-2} $ \\
 \quad $B \to D_0^* \, \ell^+ \, \nu_\ell$ & $\left(0.42 \pm 0.08\right) \times 10^{-2}$ & $\left(0.39 \pm 0.07\right) \times 10^{-2}$ \\
 \quad $B \to D_1^* \, \ell^+ \, \nu_\ell$ & $\left(0.42 \pm 0.09\right) \times 10^{-2}$ & $\left(0.39 \pm 0.08\right) \times 10^{-2}$  \\
 \quad $B \to D_1 \, \ell^+ \, \nu_\ell$ & $\left(0.66 \pm 0.11\right) \times 10^{-2}$ & $\left(0.62 \pm 0.10\right) \times 10^{-2}$ \\
 \quad $B \to D_2^* \, \ell^+ \, \nu_\ell$ & $\left(0.29 \pm 0.03 \right) \times 10^{-2}$ & $\left(0.27 \pm 0.03 \right) \times 10^{-2}$  \\
 \quad $B \to D \pi \pi \, \ell^+ \, \nu_\ell$ & $\left(0.06 \pm 0.09 \right) \times 10^{-2}$ & $\left(0.06 \pm 0.09 \right) \times 10^{-2}$  \\
 \quad $B \to D^* \pi \pi \, \ell^+ \, \nu_\ell$ & $\left(0.22 \pm 0.10 \right) \times 10^{-2}$ & $\left(0.20 \pm 0.10 \right) \times 10^{-2}$  \\
 \quad $B \to D_s K \, \ell^+ \, \nu_\ell$ & $\left(0.03 \pm 0.01 \right) \times 10^{-2}$ & -  \\
 \quad $B \to D^*_s K \, \ell^+ \, \nu_\ell$ & $\left(0.03 \pm 0.01 \right) \times 10^{-2}$ & -  \\
 \quad $B \to D \eta \, \ell^+ \, \nu_\ell$ & $\left(0.41 \pm 0.41 \right) \times 10^{-2}$ & $\left(0.41 \pm 0.41 \right) \times 10^{-2}$  \\
 \quad $B \to D^{*} \eta \, \ell^+ \, \nu_\ell$ & $\left(0.41 \pm 0.41 \right) \times 10^{-2}$ & $\left(0.41 \pm 0.41 \right) \times 10^{-2}$  \\
   \hline
    \quad $B \to X_u \, \ell^+ \, \nu_\ell$ & $\left(0.22 \pm 0.03\right) \times 10^{-2} $ & $\left(0.21 \pm 0.03 \right) \times 10^{-2} $ \\
  \hline
   $B \to X \, \ell^+ \, \nu_\ell$ & $\left(10.99 \pm 0.28\right) \times 10^{-2} $ & $\left(10.33 \pm 0.28\right) \times 10^{-2} $ \\
  
 \hline\hline
\end{tabular}
\end{table}

\section{Analysis strategy}

Our general analysis strategy is as follows: we first identify samples of same-sign and opposite-sign di-electron candidates. We then apply a selection based on tracks and global event properties to enrich the samples with double-semileptonic $B^0 - \overline{B^0}$ ("signal") decays. For events passing the selection, we construct a variable that can distinguish between signal and misreconstructed events, which we then fit to extract $N_{\mathrm{SS}}$ and $N_{\mathrm{OS}}$. We describe these steps in more detail below. 

\subsection{Di-electron candidate selection}
We form electron pairs from single electron candidates that satisfy the following criteria. We demand the single electron candidate having a center-of-mass momentum of $|p^*_e|>1$~GeV. In addition, we demand that the impact parameter is consistent with the interaction point by requiring the track to pass the interaction point within 4~cm along the beam axis in the laboratory frame and within 2~cm transverse to the beam axis. Furthermore, we require the single electron candidates to have an electron ID likelihood above 0.9. The efficiency of this selection for single electron candidates is $93.7$\% while the pion-electron misidentification rate is $2.2$\%.

Each electron candidate energy is corrected for loss due to bremsstrahlung radiation and final state radiation using a dedicated search algorithm that matches electromagnetic clusters in the calorimeter to electron candidates. We then separate the electron pair candidates according to their reconstructed charges into SS and OS categories. 

\subsection{Signal enrichment}
Using the correspondence between the generated particles and the reconstructed tracks, we classify the pair as ``Signal" if both electrons are daughters of different $B$ mesons, and label the rest ``Other". We use the off-resonance dataset to describe the continuum background distributions. 

We use the following selections to suppress the major backgrounds. First, we reject lepton pairs coming from photon conversion or resonant decay by checking whether either of the signal electron candidates can match to an opposite-sign track in the event under the electron mass hypothesis and give an invariant mass near zero ($m_{ee}<0.2$~GeV) or near the $J/\psi$ resonance ($\SI{2.92}{\gev}<m_{ee}<\SI{3.14}{\gev}$). We discard di-electron candidates if either criterion is satisfied ($\epsilon_{\textrm{SS}}=0.958$, $\epsilon_{\textrm{OS}}=0.896$). Next, if multiple di-electron pair candidates are present in an event, we randomly choose one as the best candidate in the event ($\epsilon_{\textrm{SS}}=0.999$, $\epsilon_{\textrm{OS}}=0.990$). We discard events in which there are less than five total tracks that pass the same impact parameter selections as the signal leptons ($\epsilon_{\textrm{SS}}=0.923$, $\epsilon_{\textrm{OS}}=0.917$). In order to suppress continuum backgrounds, we then retain only events with a normalized Fox-Wolfram moment~\cite{PhysRevLett.41.1581} value $R_2<0.3$ ($\epsilon_{\textrm{SS}}=0.889$, $\epsilon_{\textrm{OS}}=0.895$). Finally, we discard events for which electron ID corrections are not available for both electron candidates  ($\epsilon_{\textrm{SS}}=0.890$, $\epsilon_{\textrm{OS}}=0.894$).

After these selections, the remaining backgrounds are largely from mis-identified electrons, electron pairs in which both electrons have the same common $B$-meson ancestor, $B\to (X_c \to X_s e^- \overline{\nu_e}) e^+ \nu_e$ (OS only), and continuum events. 

\subsection{The extraction variable}
\begin{figure}[h!]
\includegraphics[width=0.45\textwidth]{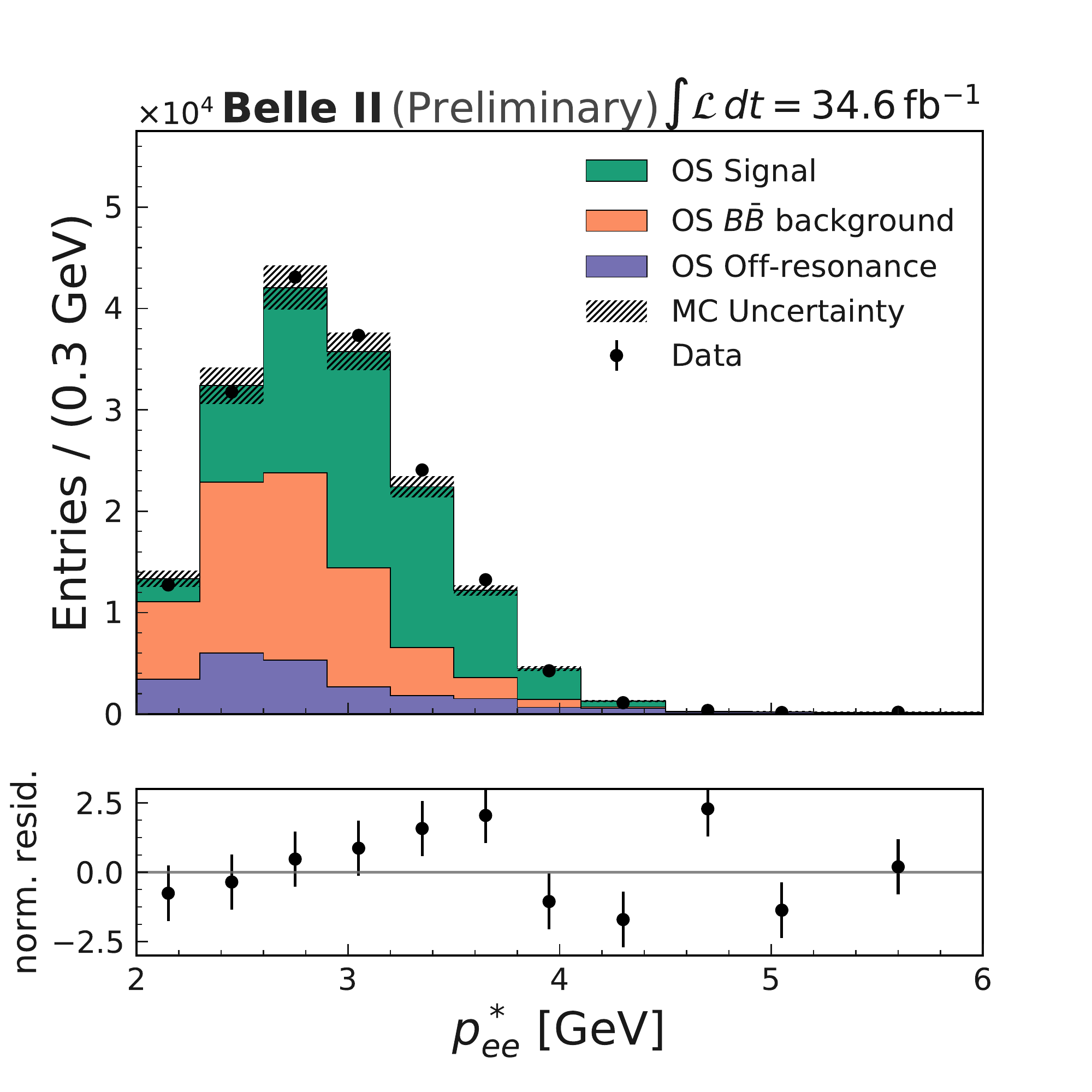}
\includegraphics[width=0.45\textwidth]{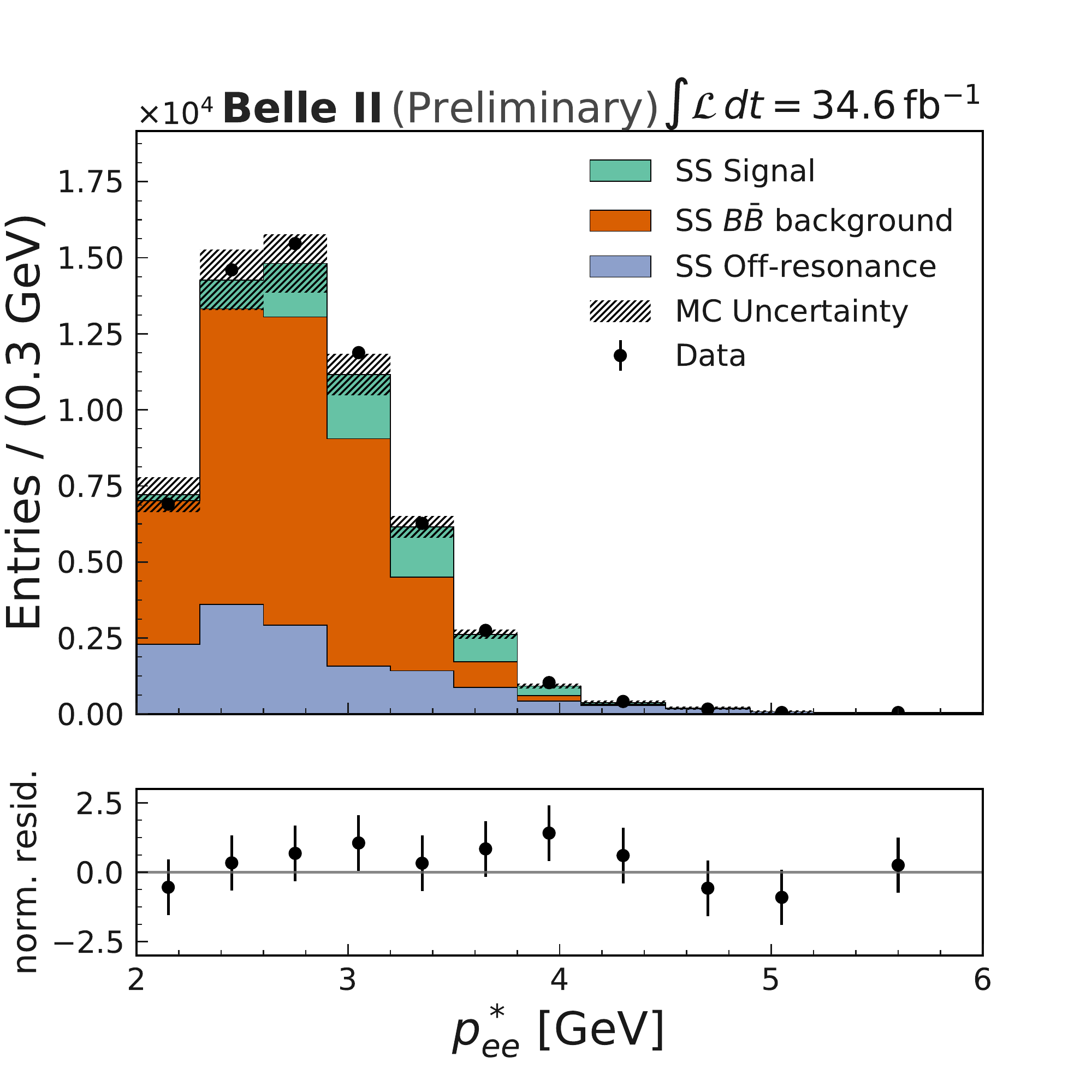}
\caption{The \pee spectrum for opposite-sign (left) and same-sign (right) di-electron samples after all selections and before fitting. The shaded, stacked histograms show the expected spectra from the sum of ``Signal" MC (green), ``Other" $B\overline{B}$ MC (orange), and scaled off-resonance data (purple). The black points with error bars indicate the spectrum as measured in data. The shaded area shows the size of the systematic uncertainty from lepton identification efficiencies, signal and background shapes, and the statistical uncertainty of the off-resonance data. The panels below the histograms show the normalized residuals between data and MC calculated as $\nicefrac{(N_\text{data}-N_\text{MC})}{\sqrt{\sigma^2_\text{data}+\sigma^2_\text{MC}}}$, where $N$ is the number of entries, $\sigma_\text{data}$ denotes the statistical uncertainty of the data and $\sigma_\text{MC}$ is the total uncertainty of MC in each bin.}
\label{fig:pll_prefit}
\end{figure}

Our signal consists of two electrons from semileptonic $B$ decays, each of which has a mean energy above that of the primary background sources. Therefore, the sum of the magnitude of the momenta in the center-of-mass frame of the two electrons:
\begin{align}
 \pee = \left| p_{e1}^* \right| + \left| p_{e2}^* \right| \, 
\end{align}
provides discrimination between signal and background. This variable has not been exploited thus far in time-integrated $\chi_d$ measurements. In Fig.~\ref{fig:pll_prefit} we show this spectrum in OS and SS for MC and data. 

\section{Fitting procedure}

The number of same-sign and opposite-sign \bxenu candidates is determined by a simultaneous binned likelihood fit to the \pee distribution of both samples and in 11 \pee bins ranging from 2 - 6 GeV. For each of the 11 \pee bins, the free parameters of the fit are the number of same-sign and opposite-sign \bxenu candidates $N_{\mathrm{SS}}$, $N_{\mathrm{OS}}$, the number of background events in each sample from $B$ meson decays $N_{\mathrm{BSS}}$,  $N_{\mathrm{BOS}}$ as well as he number of background events in each sample from continuum processes $N_{\mathrm{CSS}}$,  $N_{\mathrm{COS}}$ 

These parameters correspond to the yields of the three event categories considered.
The total likelihood function is 
\begin{align}\label{eq:likelihood}
 \mathcal{L} =  \prod_i^{\rm bins} \, \mathcal{P}\left( n_i ; \nu_i \right)  \,  \times \prod_k \, \mathcal{G}_k \, ,
\end{align}
with $n_i$ denoting the number of observed data events and $\nu_i$ the total number of expected events in a given bin $i$. Here, $\mathcal{G}_k$ are nuisance-parameter (NP) constraints for a given template $k$, whose role is to incorporate systematic uncertainties and the number of expected continuum events, as determined from the off-resonance data, directly into the fit. 
The number of expected events in a given bin, $\nu_i$, is estimated using MC and off-resonance data and is given by
\begin{align} \label{eq:fSS}
 \nu_i = N_{\mathrm{SS}} \cdot f_{i, \mathrm{SS}} + N_{\mathrm{BSS}} \cdot f_{i, \mathrm{BSS}} + N_{\mathrm{CSS}} \cdot f_{i, \mathrm{CSS}} \, ,
\end{align} 
or 
\begin{align} \label{eq:fOS}
 \nu_i = N_{\mathrm{OS}} \cdot f_{i, \mathrm{OS}} + N_{\mathrm{BOS}} \cdot f_{i, \mathrm{BOS}} + N_{\mathrm{COS}} \cdot f_{i, \mathrm{COS}}  \, ,
\end{align}
for same-sign or opposite-sign events, respectively. Here the $f_i$ correspond to the fraction of events of each category being reconstructed in bin $i$ as determined by the MC simulation or the off-resonance data. The NP constraints are constructed such that they take into account uncertainties due to the electron identification efficiency, signal and background template compositions, and the statistical uncertainty of the template in question. They are incorporated using multivariate Gaussian distributions,  $\mathcal{G}_k = \mathcal{G}_k( \boldsymbol{0}; \boldsymbol{\theta}_k, \Sigma_k ) $. Here $\Sigma_k$ denotes the systematic covariance matrix for a given template $k$ and $\boldsymbol{\theta}_k$ is a vector of NPs. The covariance matrix $\Sigma_k$  is the sum over all possible uncertainty sources for a given template. The fractions in Eqs.~\ref{eq:fSS} and \ref{eq:fOS} are allowed to change within these systematic uncertainties according to:
\begin{equation}
 f_i = \frac{ N_{i}^{\rm MC} \left( 1 + \theta_{i} \right) }{ \sum_j N_{j}^{\rm MC} \left( 1 + \theta_{j} \right)  } \, ,
\end{equation}
with $N_{i}^{\rm MC}$ denoting the number of expected events of a given category in bin $i$ as estimated by MC, and $\theta_{j}$ the $j^\text{th}$ element of the NP vector $\boldsymbol{\theta}_k$. The likelihood function is maximized numerically to determine all components and NP constraints. Confidence intervals for the $N_{\mathrm{SS}}$ and $N_{\mathrm{OS}}$ components are constructed using the profile likelihood ratio method.

\section{Results}

\begin{figure}[h!]
\includegraphics[width=0.45\textwidth]{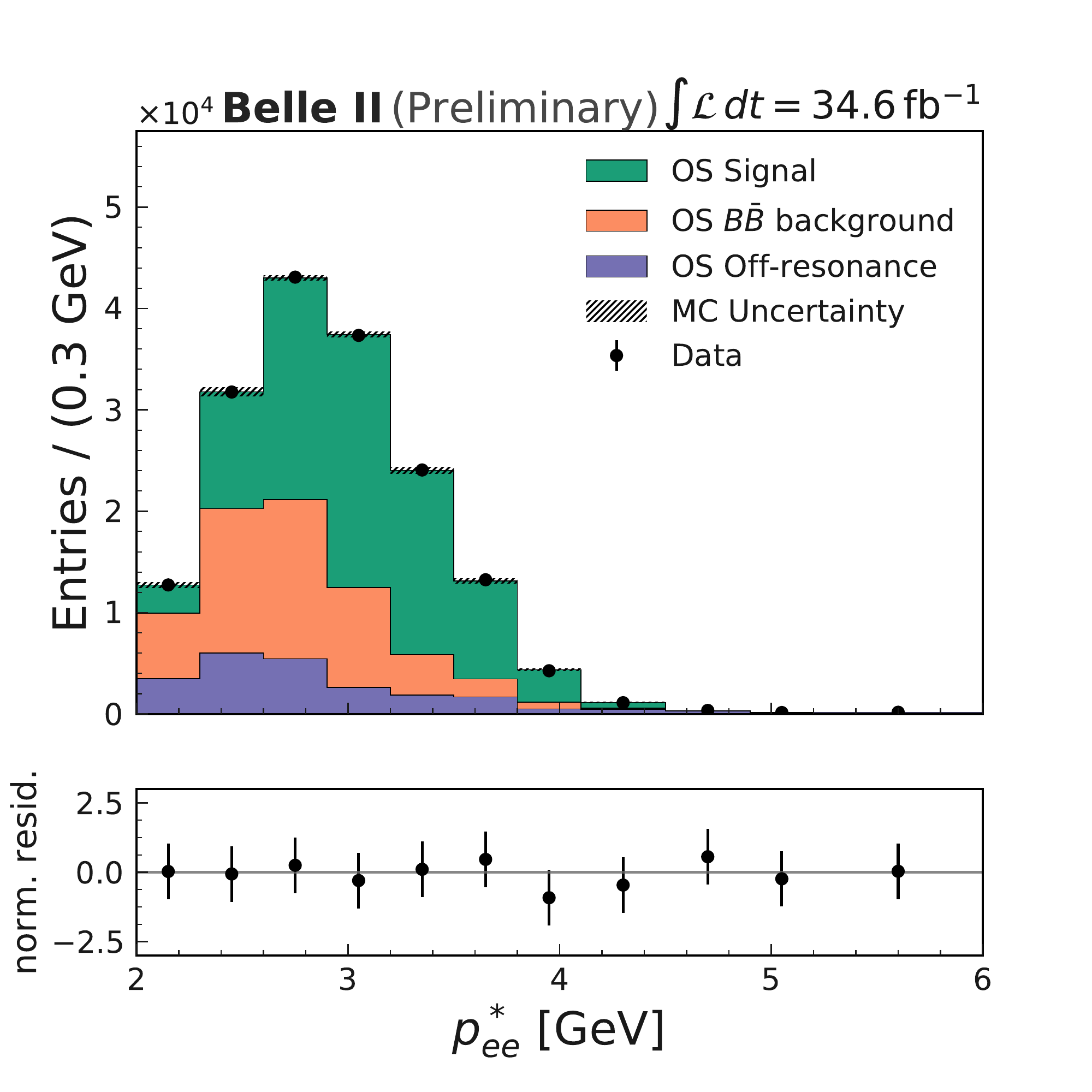}
\includegraphics[width=0.45\textwidth]{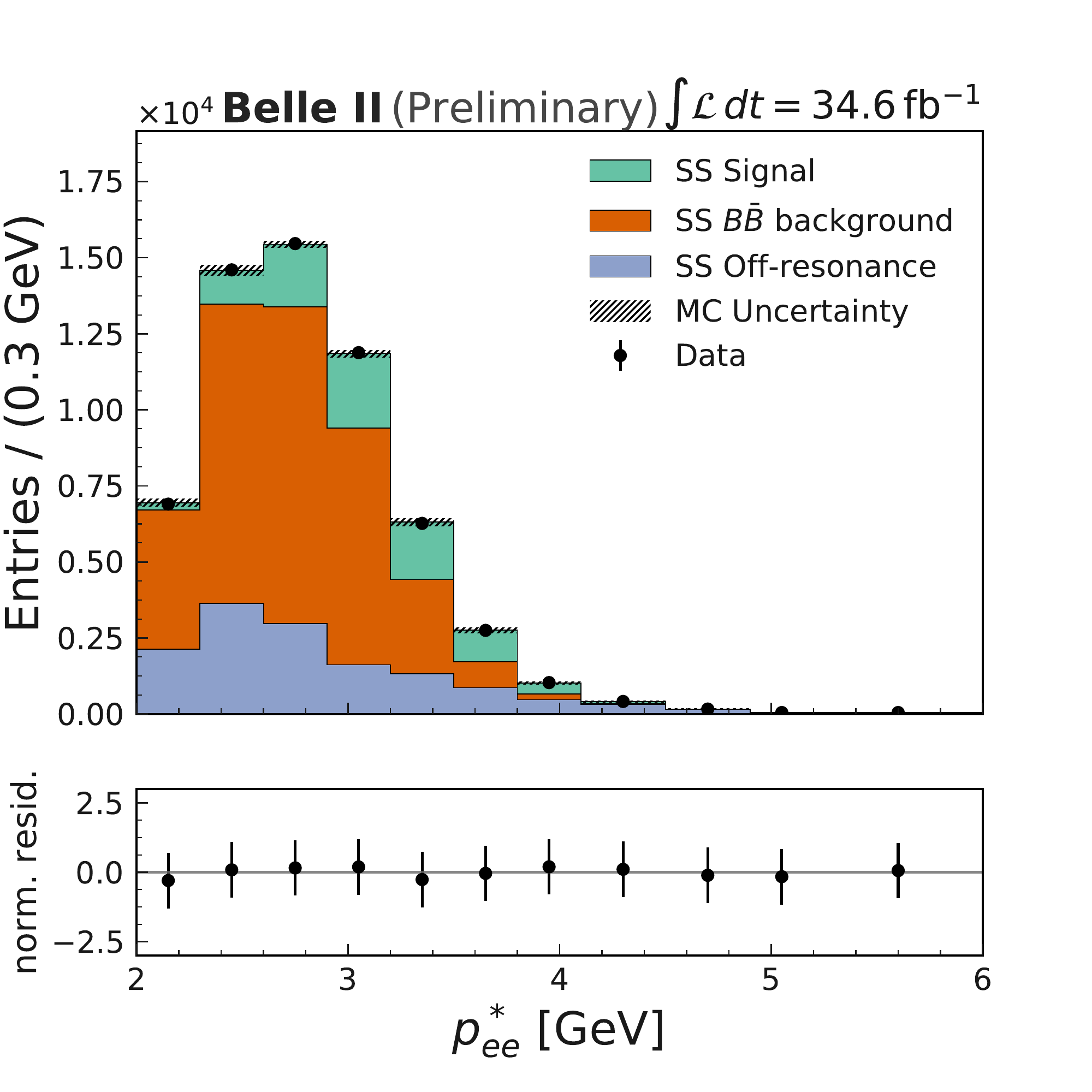}
\caption{The \pee spectrum for opposite-sign (left) and same-sign (right) di-electron samples after fitting. The shaded, stacked histograms show the fitted spectra from the sum of ``Signal" MC (green), ``Other" $B\overline{B}$ MC (orange), and scaled off-resonance data (purple). The black points with error bars indicate the spectrum as measured in data. The shaded area around the MC expectation correspond to the post-fit uncertainty. The panels below the histograms show the normalized residuals between data and MC calculated as $\nicefrac{(N_\text{data}-N_\text{MC})}{\sqrt{\sigma^2_\text{data}+\sigma^2_\text{MC}}}$, where $N$ is the number of entries, $\sigma_\text{data}$ denotes the statistical uncertainty of the data and $\sigma_\text{MC}$ is the total uncertainty of MC in each bin..}
\label{fig:pll_postfit}
\end{figure}

Figure~\ref{fig:pll_postfit} shows the post-fit \pee distribution for same-sign and opposite-sign candidates. The goodness-of-fit cannot be judged based on the residuals alone, but also depends on the NP pulls, which we show in Appendix~\ref{sec:appendix}. From this fit, we measure:
\begin{align} \label{eq:chi_d_res}
\chi_d =  \chidmeas \, 
\end{align}
 where the first uncertainty is statistical and the second is systematic. 
 The overall uncertainty is dominated by the statistical uncertainty of the same-sign sample, the limited size of the off-resonance data sample, and the systematic uncertainty of the electron identification corrections. The size of these systematic uncertainties is expected to decrease with larger control samples. The obtained value of $\chi_d$ is compatible with the world average from time-independent and time-dependent determinations~\cite{pdg:2020} and already has a precision comparable to the time-independent world average.
   
\newpage
\section{ACKNOWLEDGEMENTS}
We thank the SuperKEKB group for the excellent operation of the
accelerator; the KEK cryogenics group for the efficient
operation of the solenoid; and the KEK computer group for
on-site computing support.
This work was supported by the following funding sources:
Science Committee of the Republic of Armenia Grant No. 18T-1C180;
Australian Research Council and research grant Nos.
DP180102629, 
DP170102389, 
DP170102204, 
DP150103061, 
FT130100303, 
and
FT130100018; 
Austrian Federal Ministry of Education, Science and Research, and
Austrian Science Fund No. P 31361-N36; 
Natural Sciences and Engineering Research Council of Canada, Compute Canada and CANARIE;
Chinese Academy of Sciences and research grant No. QYZDJ-SSW-SLH011,
National Natural Science Foundation of China and research grant Nos.
11521505,
11575017,
11675166,
11761141009,
11705209,
and
11975076,
LiaoNing Revitalization Talents Program under contract No. XLYC1807135,
Shanghai Municipal Science and Technology Committee under contract No. 19ZR1403000,
Shanghai Pujiang Program under Grant No. 18PJ1401000,
and the CAS Center for Excellence in Particle Physics (CCEPP);
the Ministry of Education, Youth and Sports of the Czech Republic under Contract No.~LTT17020 and 
Charles University grants SVV 260448 and GAUK 404316;
European Research Council, 7th Framework PIEF-GA-2013-622527, 
Horizon 2020 Marie Sklodowska-Curie grant agreement No. 700525 `NIOBE,' 
and
Horizon 2020 Marie Sklodowska-Curie RISE project JENNIFER2 grant agreement No. 822070 (European grants);
L'Institut National de Physique Nucl\'{e}aire et de Physique des Particules (IN2P3) du CNRS (France);
BMBF, DFG, HGF, MPG, AvH Foundation, and Deutsche Forschungsgemeinschaft (DFG) under Germany's Excellence Strategy -- EXC2121 ``Quantum Universe''' -- 390833306 (Germany);
Department of Atomic Energy and Department of Science and Technology (India);
Israel Science Foundation grant No. 2476/17
and
United States-Israel Binational Science Foundation grant No. 2016113;
Istituto Nazionale di Fisica Nucleare and the research grants BELLE2;
Japan Society for the Promotion of Science,  Grant-in-Aid for Scientific Research grant Nos.
16H03968, 
16H03993, 
16H06492,
16K05323, 
17H01133, 
17H05405, 
18K03621, 
18H03710, 
18H05226,
19H00682, 
26220706,
and
26400255,
the National Institute of Informatics, and Science Information NETwork 5 (SINET5), 
and
the Ministry of Education, Culture, Sports, Science, and Technology (MEXT) of Japan;  
National Research Foundation (NRF) of Korea Grant Nos.
2016R1\-D1A1B\-01010135,
2016R1\-D1A1B\-02012900,
2018R1\-A2B\-3003643,
2018R1\-A6A1A\-06024970,
2018R1\-D1A1B\-07047294,
2019K1\-A3A7A\-09033840,
and
2019R1\-I1A3A\-01058933,
Radiation Science Research Institute,
Foreign Large-size Research Facility Application Supporting project,
the Global Science Experimental Data Hub Center of the Korea Institute of Science and Technology Information
and
KREONET/GLORIAD;
Universiti Malaya RU grant, Akademi Sains Malaysia and Ministry of Education Malaysia;
Frontiers of Science Program contracts
FOINS-296,
CB-221329,
CB-236394,
CB-254409,
and
CB-180023, and SEP-CINVESTAV research grant 237 (Mexico);
the Polish Ministry of Science and Higher Education and the National Science Center;
the Ministry of Science and Higher Education of the Russian Federation,
Agreement 14.W03.31.0026;
University of Tabuk research grants
S-1440-0321, S-0256-1438, and S-0280-1439 (Saudi Arabia);
Slovenian Research Agency and research grant Nos.
J1-9124
and
P1-0135; 
Agencia Estatal de Investigacion, Spain grant Nos.
FPA2014-55613-P
and
FPA2017-84445-P,
and
CIDEGENT/2018/020 of Generalitat Valenciana;
Ministry of Science and Technology and research grant Nos.
MOST106-2112-M-002-005-MY3
and
MOST107-2119-M-002-035-MY3, 
and the Ministry of Education (Taiwan);
Thailand Center of Excellence in Physics;
TUBITAK ULAKBIM (Turkey);
Ministry of Education and Science of Ukraine;
the US National Science Foundation and research grant Nos.
PHY-1807007 
and
PHY-1913789, 
and the US Department of Energy and research grant Nos.
DE-AC06-76RLO1830, 
DE-SC0007983, 
DE-SC0009824, 
DE-SC0009973, 
DE-SC0010073, 
DE-SC0010118, 
DE-SC0010504, 
DE-SC0011784, 
DE-SC0012704; 
and
the National Foundation for Science and Technology Development (NAFOSTED) 
of Vietnam under contract No 103.99-2018.45.

\bibliography{belle2}

\providecommand{\href}[2]{#2}\begingroup\raggedright\begin{thebibliography}{10}

\bibitem{Prentice:1987ap}
H.~Albrecht et al., {\em Observation of $B^0$-$\bar{B}^0$ mixing\/},
  \href{http://dx.doi.org/https://doi.org/10.1016/0370-2693(87)91177-4}{Physics
  Letters B {\bf 192} (1987) no.~1, 51--63}.

\bibitem{pdg:2020}
P.~Zyla et al., {Particle Data Group} Prog. Theor. Exp. Phys. {\bf 2020} {\bf
  083C01} (2020)  .

\bibitem{Abe:2010sj}
T.~Abe et al., {Belle II Collaboration}, {\em {Belle II Technical Design
  Report}\/},  \href{http://arxiv.org/abs/1011.0352}{{\tt arXiv:1011.0352
  [physics.ins-det]}}.

\bibitem{ref:b2tip}
E.~Kou et al., {\em The Belle II Physics Book\/},
  \href{http://dx.doi.org/10.1093/ptep/ptz106}{PTEP {\bf 2019} (2019) no.~12,
  123C01}.

\bibitem{superkekb}
K.~Akai, K.~Furukawa, and H.~Koiso, {SuperKEKB Collaboration}, {\em {SuperKEKB
  Collider}\/},  \href{http://dx.doi.org/10.1016/j.nima.2018.08.017}{Nucl.
  Instrum. Meth. {\bf A907} (2018)  188--199}.

\bibitem{Lange:2001uf}
D.~J. Lange, {\em {The EvtGen particle decay simulation package}\/},
\href{http://dx.doi.org/10.1016/S0168-9002(01)00089-4}{Nucl. Instrum. Meth.
  {\bf A462} (2001)  152--155}.

\bibitem{Boyd:1994tt}
C.~G. Boyd, B.~Grinstein, and R.~F. Lebed, {\em {Constraints on form-factors
  for exclusive semileptonic heavy to light meson decays}\/},
  \href{http://dx.doi.org/10.1103/PhysRevLett.74.4603}{Phys. Rev. Lett. {\bf
  74} (1995)  4603--4606},
\href{http://arxiv.org/abs/hep-ph/9412324}{{\tt arXiv:hep-ph/9412324
  [hep-ph]}}.

\bibitem{Glattauer:2015teq}
R.~Glattauer et al., {Belle Collaboration}, {\em {Measurement of the decay
  $B\to D\ell\nu_\ell$ in fully reconstructed events and determination of the
  Cabibbo-Kobayashi-Maskawa matrix element $|V_{cb}|$}\/},
  \href{http://dx.doi.org/10.1103/PhysRevD.93.032006}{Phys. Rev. D {\bf 93}
  (2016) no.~3, 032006}, \href{http://arxiv.org/abs/1510.03657}{{\tt
  arXiv:1510.03657 [hep-ex]}}.

\bibitem{Grinstein:2017nlq}
B.~Grinstein and A.~Kobach, {\em {Model-Independent Extraction of $|V_{cb}|$
  from $\bar{B}\rightarrow D^* \ell \overline{\nu}$}\/},
  \href{http://dx.doi.org/10.1016/j.physletb.2017.05.078}{Phys. Lett. {\bf B
  771} (2017)  359--364},
\href{http://arxiv.org/abs/1703.08170}{{\tt arXiv:1703.08170 [hep-ph]}}.

\bibitem{Bigi:2017njr}
D.~Bigi, P.~Gambino, and S.~Schacht, {\em {A fresh look at the determination of
  $|V_{cb}|$ from $B\to D^{*} \ell \nu$}\/},
  \href{http://dx.doi.org/10.1016/j.physletb.2017.04.022}{Phys. Lett. B {\bf
  769} (2017)  441--445}, \href{http://arxiv.org/abs/1703.06124}{{\tt
  arXiv:1703.06124 [hep-ph]}}.

\bibitem{Waheed:2018djm}
E.~Waheed et al., {Belle Collaboration}, {\em {Measurement of the CKM matrix
  element $|V_{cb}|$ from $B^0\to D^{*-}\ell^ {+} \nu_\ell$ at Belle}\/},
  \href{http://dx.doi.org/10.1103/PhysRevD.100.052007}{Phys. Rev. D {\bf 100}
  (2019) no.~5, 052007}, \href{http://arxiv.org/abs/1809.03290}{{\tt
  arXiv:1809.03290 [hep-ex]}}.

\bibitem{Amhis:2019ckw}
Y.~S. Amhis et al., {HFLAV}, {\em {Averages of $b$-hadron, $c$-hadron, and
  $\tau$-lepton properties as of 2018}\/},
  \href{http://arxiv.org/abs/1909.12524}{{\tt arXiv:1909.12524 [hep-ex]}}.

\bibitem{Bernlochner:2016bci}
F.~U. Bernlochner and Z.~Ligeti, {\em {Semileptonic $B_{(s)}$ decays to excited
  charmed mesons with $e,\mu,\tau$ and searching for new physics with
  $R(D^{**})$}\/},  \href{http://dx.doi.org/10.1103/PhysRevD.95.014022}{Phys.
  Rev. D {\bf 95} (2017) no.~1, 014022},
  \href{http://arxiv.org/abs/1606.09300}{{\tt arXiv:1606.09300 [hep-ph]}}.

\bibitem{Bernlochner:2017jxt}
F.~U. Bernlochner, Z.~Ligeti, and D.~J. Robinson, {\em {Model independent
  analysis of semileptonic $B$ decays to $D^{**}$ for arbitrary new
  physics}\/},  \href{http://dx.doi.org/10.1103/PhysRevD.97.075011}{Phys. Rev.
  D {\bf 97} (2018) no.~7, 075011}, \href{http://arxiv.org/abs/1711.03110}{{\tt
  arXiv:1711.03110 [hep-ph]}}.

\bibitem{PhysRevD.41.1496}
C.~Ramirez, J.~F. Donoghue, and G.~Burdman,
  \href{http://dx.doi.org/10.1103/PhysRevD.41.1496}{{\em Semileptonic
  $b\ensuremath{\rightarrow}u$ decay\/}, Phys. Rev. D {\bf 41} (Mar, 1990)
  1496--1503}. \url{https://link.aps.org/doi/10.1103/PhysRevD.41.1496}.

\bibitem{PhysRevLett.41.1581}
G.~C. Fox and S.~Wolfram,
  \href{http://dx.doi.org/10.1103/PhysRevLett.41.1581}{{\em Observables for the
  Analysis of Event Shapes in ${e}^{+}{e}^{\ensuremath{-}}$ Annihilation and
  Other Processes\/}, Phys. Rev. Lett. {\bf 41} (Dec, 1978)  1581--1585}.
  \url{https://link.aps.org/doi/10.1103/PhysRevLett.41.1581}.

\end{thebibliography}\endgroup
\bibliographystyle{belle2-note}

\clearpage

\appendix
\section{Nuisance parameter pulls}\label{sec:appendix}
\begin{figure}[h!]
\includegraphics[width=0.4\textwidth]{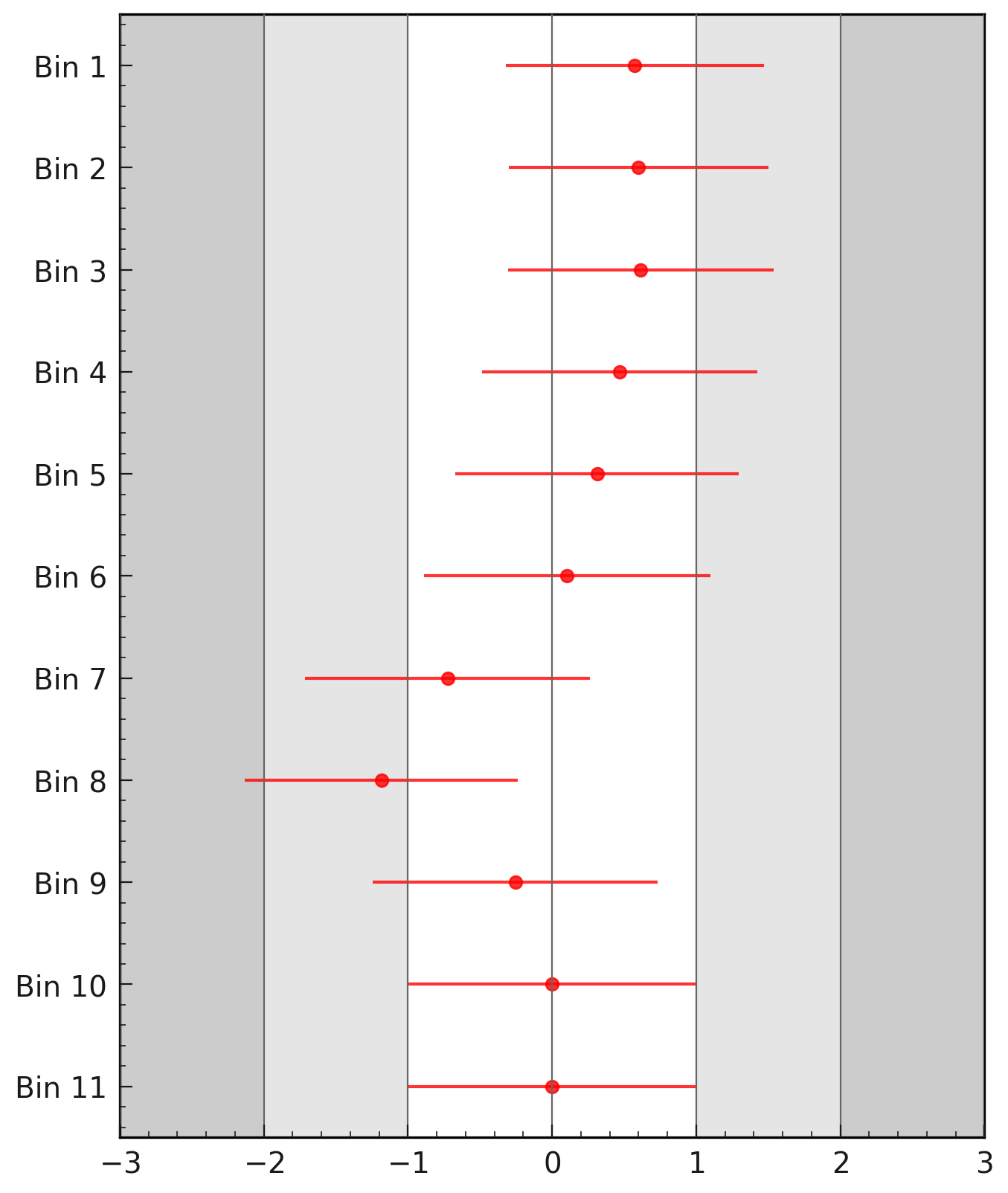}
\includegraphics[width=0.4\textwidth]{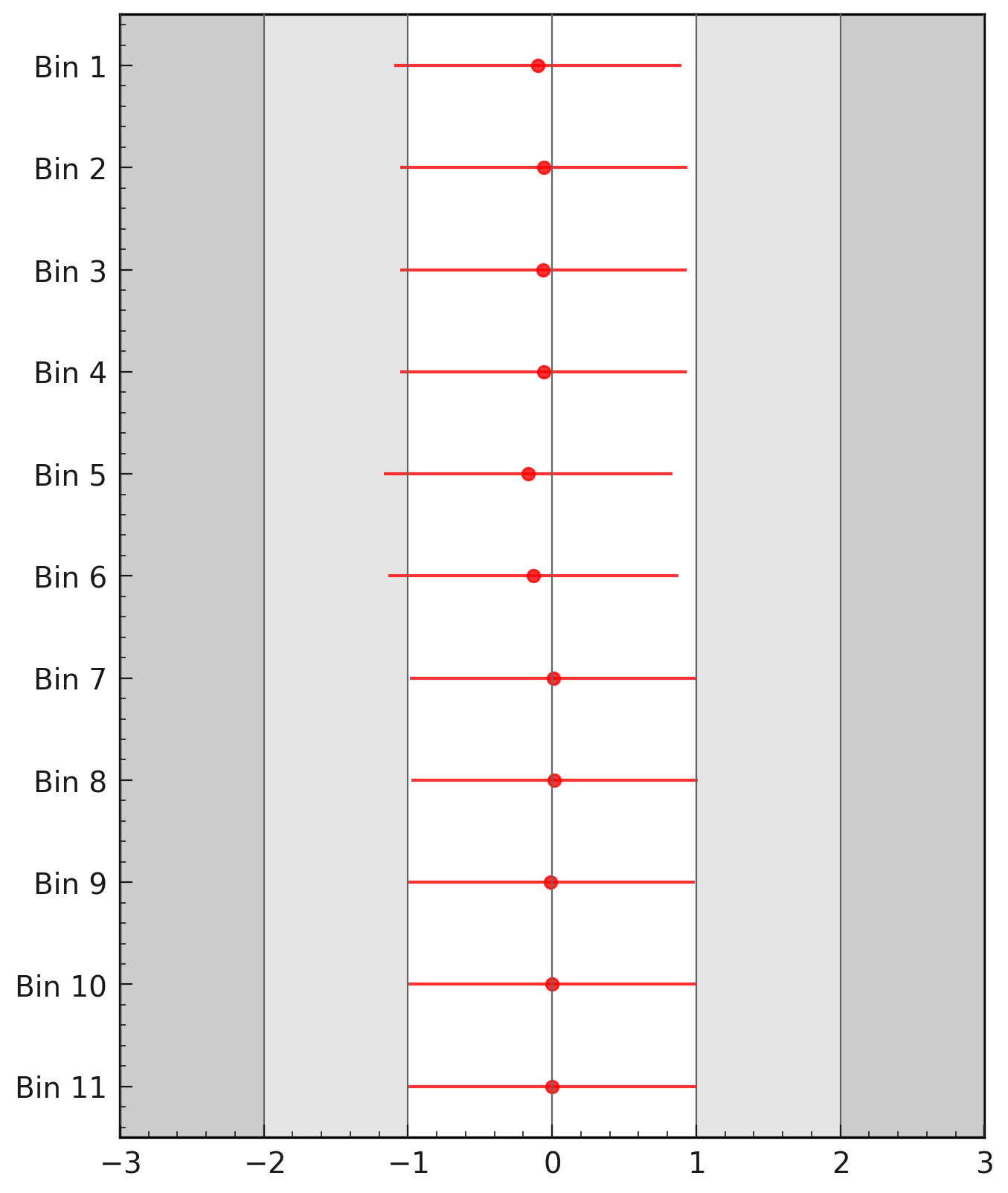}
\caption{Pulls on the nuisance parameters for the OS (left) and SS (right) signal templates.}
\label{fig:np_pulls_signal}
\end{figure}

\begin{figure}[h!]
\includegraphics[width=0.4\textwidth]{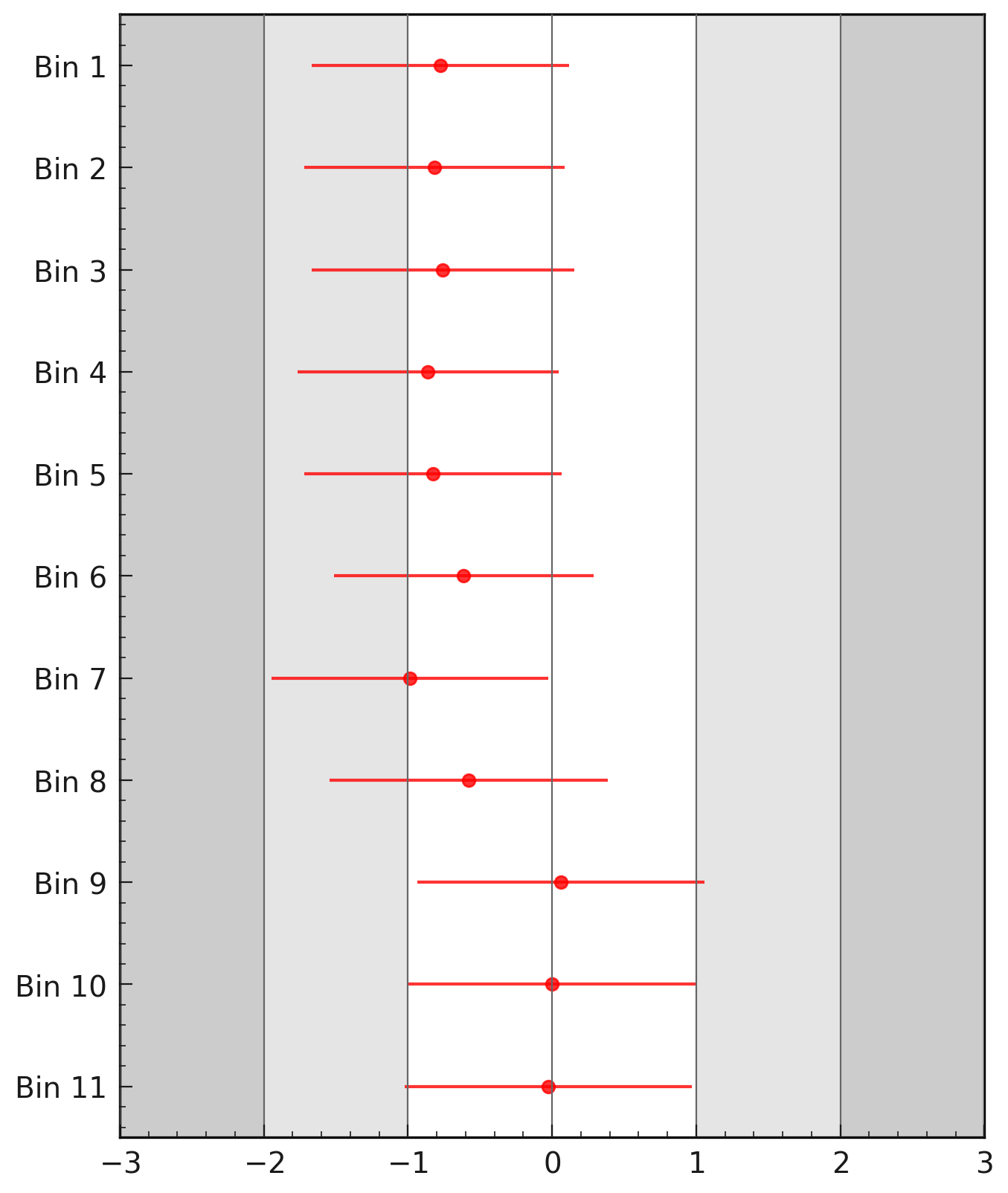}
\includegraphics[width=0.4\textwidth]{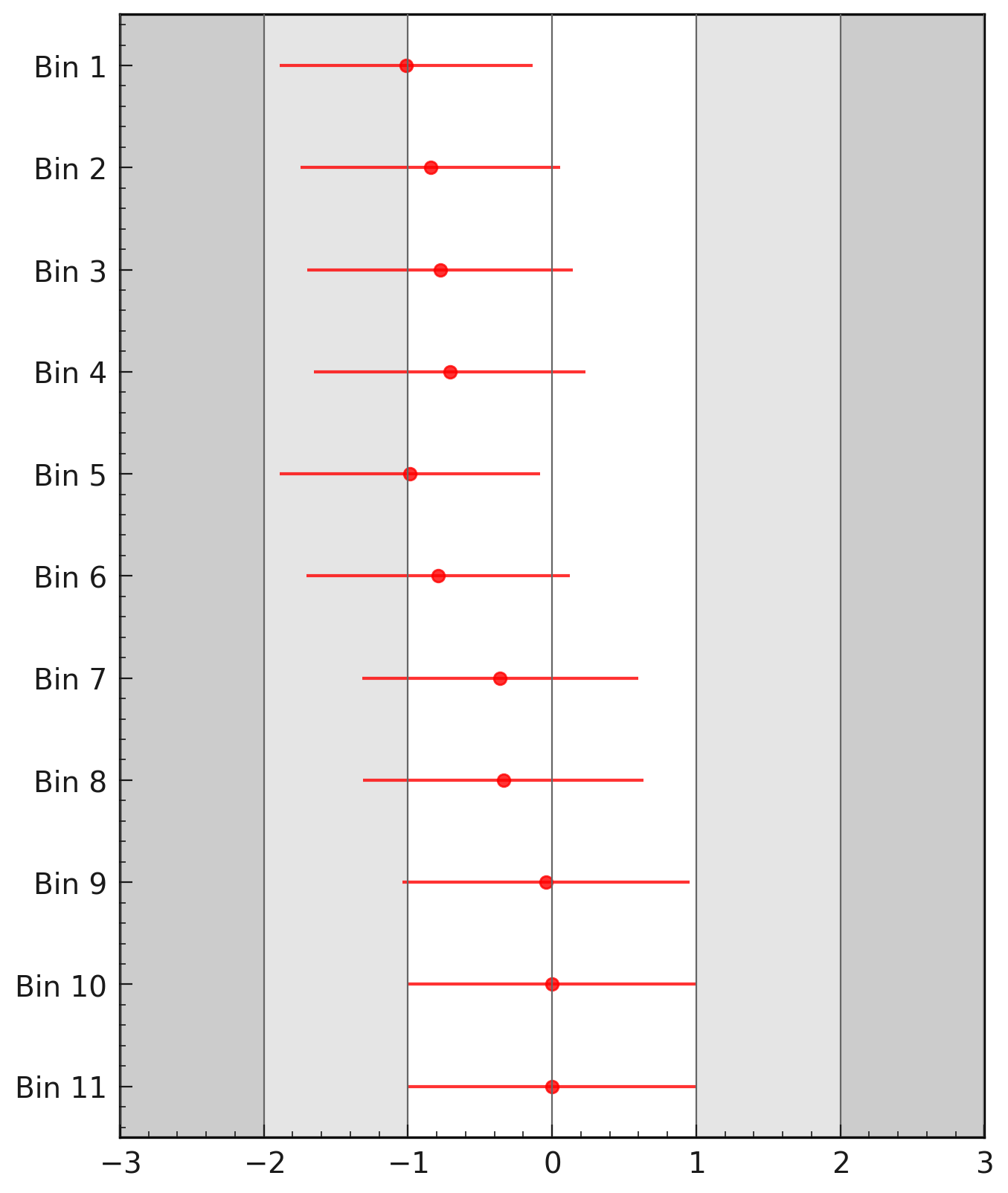}
\caption{Pulls on the nuisance parameters for the OS (left) and SS (right) $B\overline{B}$ background templates.}
\label{fig:np_pulls_bkg}
\end{figure}

\begin{figure}[h!]
\includegraphics[width=0.4\textwidth]{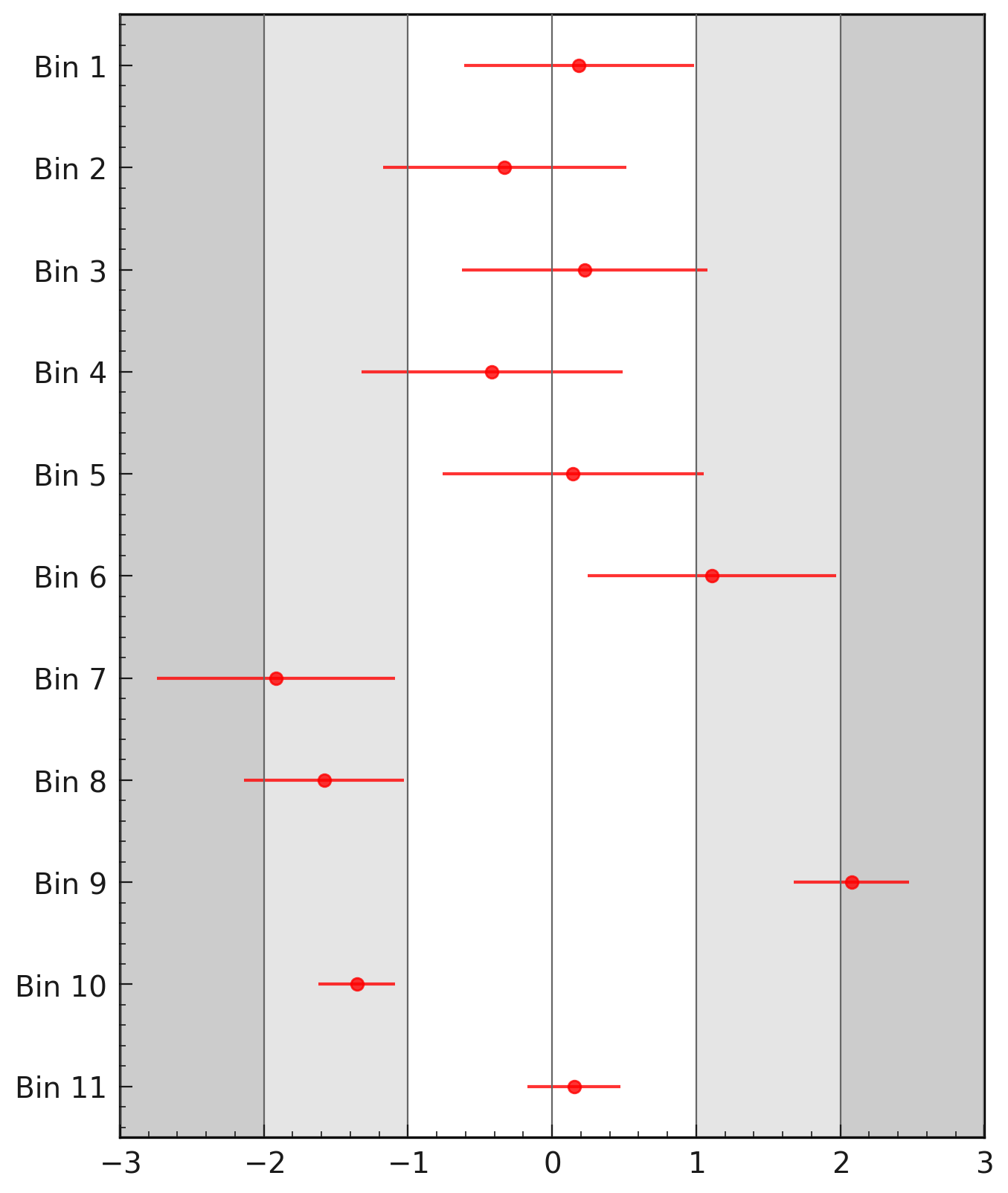}
\includegraphics[width=0.4\textwidth]{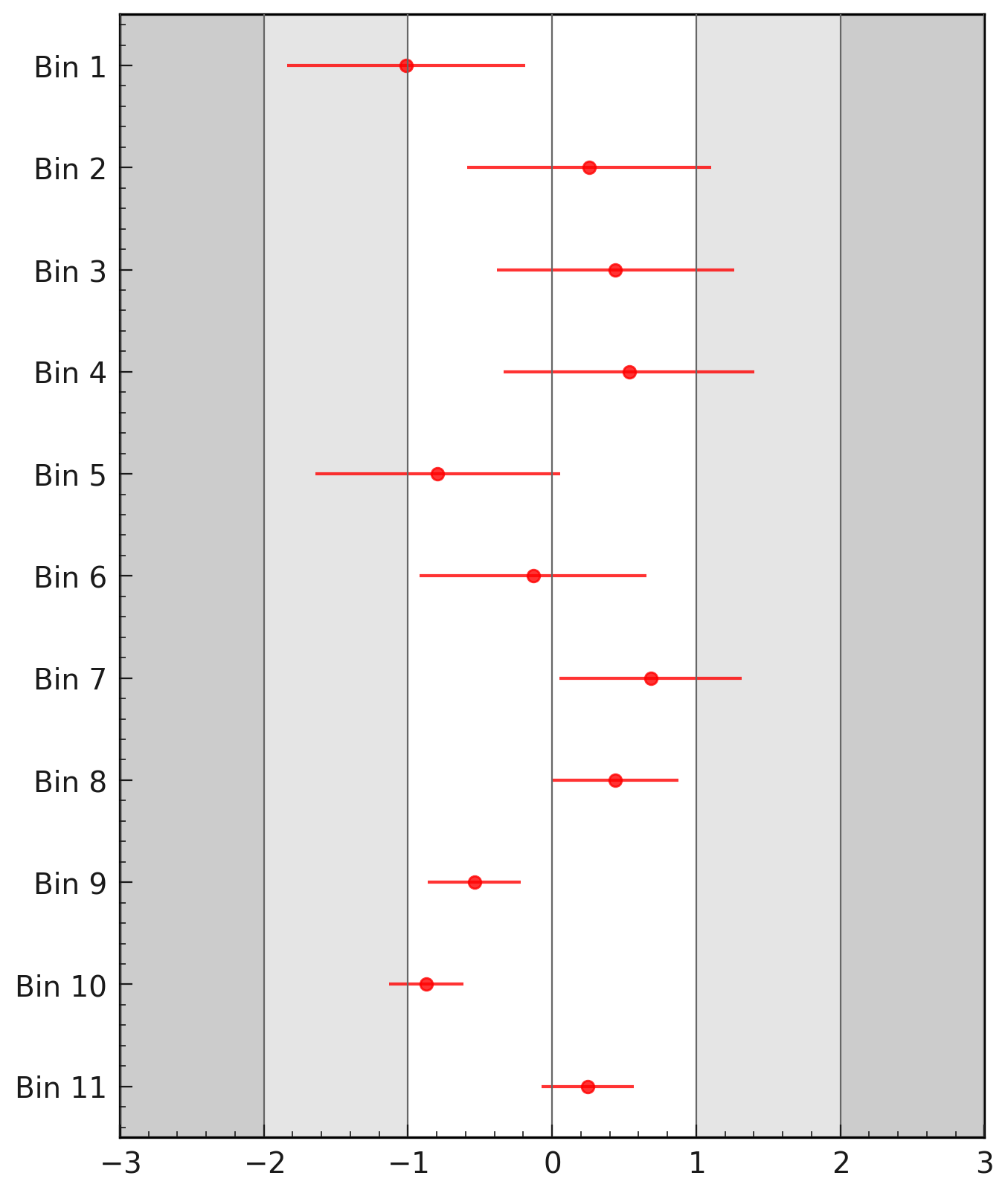}
\caption{Pulls on the nuisance parameters for the OS (left) and SS (right) off-resonance templates.}
\label{fig:np_pulls_offres}
\end{figure}

We include distributions of the NP pulls from the fit shown in Fig.~\ref{fig:pll_postfit}, for the signal (Fig.~\ref{fig:np_pulls_signal}), other $B\overline{B}$ (Fig.~\ref{fig:np_pulls_bkg}), and off-resonance (Fig.~\ref{fig:np_pulls_offres}) templates. The modest deviation that can be seen in the pulls of the OS signal template corresponds to higher values in \pee, which is mostly occupied by \bulnu decays.

\end{document}